\documentclass{PoS}

\title{A proposed very high energy electron--proton collider, VHEeP}

\ShortTitle{VHEeP: A very high energy electron--proton collider}

\author{\speaker{M. Wing}\thanks{Also supported by DESY and the Alexander von Humboldt Foundation.}\\
        UCL, London, UK\\
        E-mail: \email{m.wing@ucl.ac.uk}}

\author{A. Caldwell\\
        Max Planck Institute for Physics, Munich, Germany\\
        E-mail: \email{caldwell@mpp.mpg.de}}

\abstract{The possibility of using plasma wakefield acceleration to build a 
very high energy electron--proton (VHEeP) collider at a centre-of-mass energy of 9 TeV was
presented at the DIS2015 workshop.  In this talk, the physics case was
further developed and the idea has since been published as a journal paper.  
A brief summary is here given along with some details of the technical 
aspects not covered in the paper, which focused on the physics motivation.
It is demonstrated that an $ep$ collider
with a centre-of-mass energy a factor of 30 above HERA has sensitivity to new physical
phenomena.
}

\FullConference{XXIV International Workshop on Deep-Inelastic Scattering and Related Subjects\\
		11-15 April, 2016\\
		DESY Hamburg, Germany}

\begin{document}

\section{Introduction}

The HERA electron--proton accelerator was the first and so far only lepton--hadron collider worldwide.  With 
its centre-of-mass energy of about 300\,GeV, HERA dramatically extended the kinematic 
reach~\cite{1506.06042} for the deep inelastic scattering process compared to fixed-target experiments.  
The LHeC project~\cite{lhec} is a proposed $ep$ collider with significantly higher energy and luminosity than 
HERA with a programme to investigate Higgs physics and QCD, to search for new physics, etc..  This will use 
significant parts of the LHC infrastructure at CERN with different configurations, such as $eA$, also possible.  
In these proceedings, a very high energy electron--proton collider (VHEeP) is considered with an 
$ep$ centre-of-mass energy of about 9\,TeV, a factor of six higher than proposed for the LHeC and a factor of 30 
higher than HERA.

The VHEeP machine would strongly rely on the use of the LHC beams and the technique of plasma wakefield 
acceleration to accelerate electrons to 3\,TeV over relatively short distances.  With such an increase in centre-of-mass 
energy, the VHEeP collider will probe a new regime in deep inelastic scattering and QCD in general.  The kinematic 
regime accessible will be extended by three orders of magnitude compared to that measured at HERA.  First ideas 
on a physics case for the VHEeP collider have been outlined recently~\cite{epj:c76:463}.  These proceedings briefly 
review this as well as discussing some technical aspects of the design not covered in the publication of the physics case.  
Complementary studies of high energy $ep$ colliders have been performed elsewhere~\cite{kaya1,xia,kaya2}, 
considering both the accelerator design~\cite{kaya1,xia} and physics potential~\cite{kaya2}.

\section{VHEeP accelerator complex}

Tajima and Dawson first proposed that plasmas can sustain very large electric fields capable of accelerating bunches of 
particles~\cite{prl:43:267}.  This acceleration concept can make use of bunches of protons~\cite{pdpwa}, given  
high energy proton bunches are available and hence the possibility to have the acceleration performed in 
one stage.  Simulation has shown that the plasma wakefield created by the LHC proton bunches can 
accelerate a trailing bunch of electrons to 6\,TeV in 10\,km~\cite{pp:18:103101}.  The concept of proton-driven 
plasma wakefield acceleration will be tested by the AWAKE collaboration at CERN which aims to demonstrate 
the scheme for the first time~\cite{awake}.  The initial aims of the AWAKE experiment are to demonstrate GV/m 
accelerating gradients in plasma~\cite{awake}.  Following this, the AWAKE collaboration proposes to 
accelerate bunches of electrons to 10\,GeV in about 10\,m of plasma and demonstrate its scalability~\cite{awake-erik-ipac}.

\begin{figure}[h]
\begin{center}
\includegraphics[trim={7.5cm 3cm 11.cm 4cm},clip,width=0.5\textwidth]{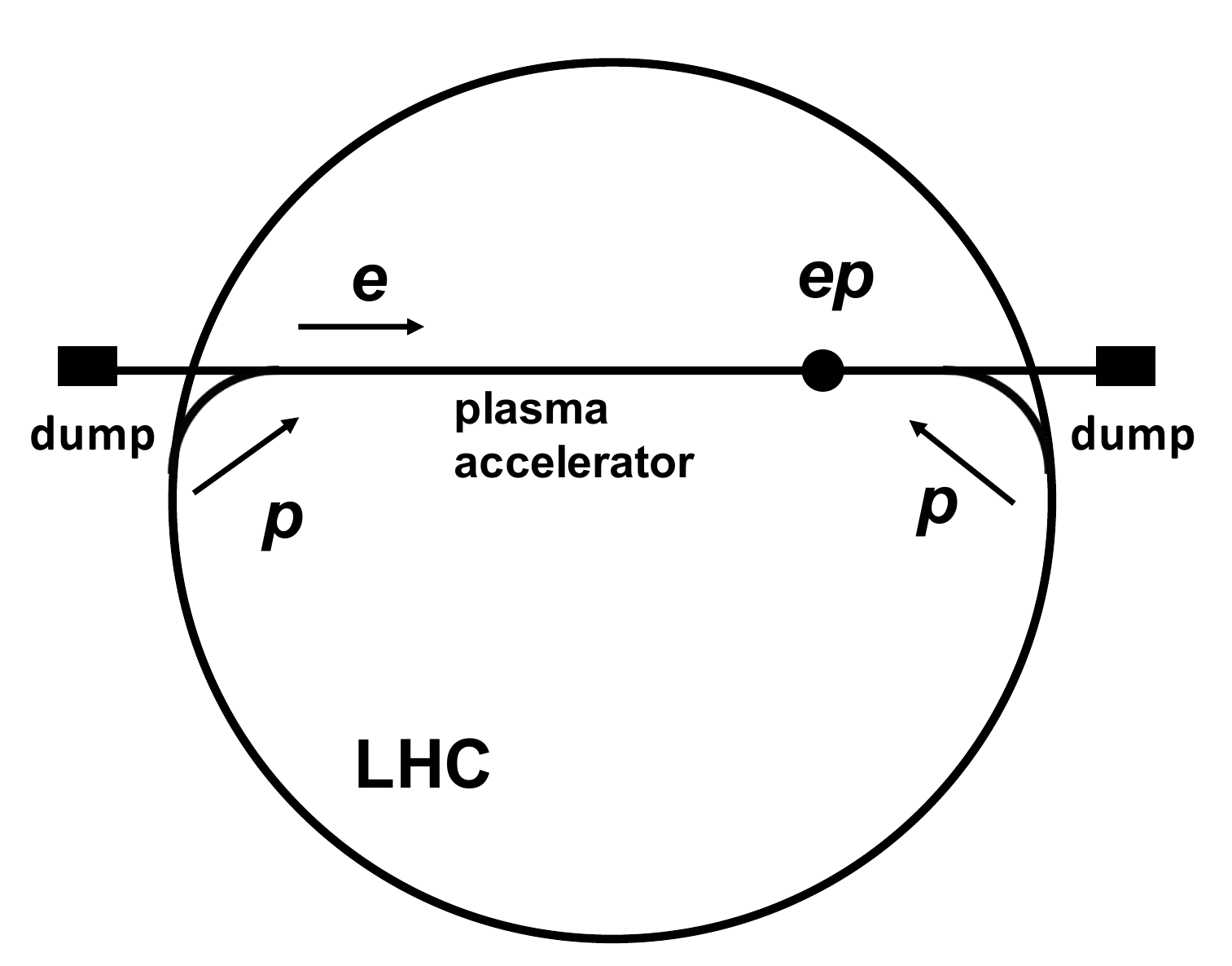}
\caption{Simple schematic of the VHEeP accelerator complex, showing the LHC ring.  Protons from the LHC are 
extracted into the VHEeP plasma accelerator and used to accelerate bunches of electrons.  Proton bunches 
rotating in the other direction in the LHC are extracted into the VHEeP tunnel and collided with electrons.  
Both proton and electron dumps could be used for fixed-target or beam-dump experiments.
}
\label{fig:acc}
\end{center}
\end{figure}

The very high energy electron--proton collider is based on current LHC infrastructure and a new tunnel to 
house the plasma accelerator.  The facility uses one 
of the LHC proton beams to generate wakefields and accelerate a trailing electron bunch which then collides with 
the other proton beam.  This is shown in a simple schematic in Fig.~\ref{fig:acc} in which the electron beam is 
chosen to have an energy of 3\,TeV, achieved in a plasma accelerator of $\lesssim$\,4\,km, and the proton beams have an 
energy of 7\,TeV.  Separation of the drive proton beam and witness electron beam will be needed to avoid $pp$ collisions; 
as well as the temporal difference, the beams will need to be separated transversely by $O$(mm).  Further study of this 
important issue is needed.

Although the energy of VHEeP will be very high, obtaining high luminosities will be, as with all plasma wakefield acceleration 
schemes, a challenge.  The luminosity, $\mathcal{L}$,  is calculated using the standard formula for a collider, 

\begin{equation}
\mathcal{L} = \frac{f \cdot N_e \cdot N_p}{4\,\pi\,\sigma_x \cdot \sigma_y} \, ,
\label{eq:lumi}
\end{equation}
where it is assumed that the number of protons in a bunch, $N_p$, is $4 \times 10^{11}$, the number of electrons in a bunch is 
$1 \times 10^{11}$ and the cross section of the bunches is dominated by the size of the proton beam which is taken to be 
$\sigma_x \sim \sigma_y = $\,4\,$\mu$m.  A frequency, $f$, of 2\,Hz is assumed given 3\,000\,bunches every 
30\,min~\cite{private:elena}.  Inserting these values into Eq.~\ref{eq:lumi}, gives 
$\mathcal{L} \sim 4 \times 10^{28}$\,cm$^{-2}$s$^{-1}$.  Running for about 80\% of the year would enable a yearly integrated 
luminosity of 1\,pb$^{-1}$ to be accumulated.

In these initial studies, an integrated luminosity over the lifetime of VHEeP of $10-100$\,pb$^{-1}$ is considered, based on the 
expected capabilities of the LHC and pre-accelerators.  The lower limit will be sufficient for measurements at low $x$ where the 
cross section is expected to rise with decreasing $x$.  A higher integrated luminosity will aid the search for physics beyond the 
Standard Model, typically at high $Q^2$.

Achieving higher luminosities than the 1\,pb$^{-1}$ per year will be a challenge.  As the cross section will be dominated 
by the proton bunch size, this could be investigated.  Another area where, at least superficially, significant gains could be made is 
the 30-min refill time of the LHC.  Studies are ongoing as part of the LHC upgrade and FCC proposal, to which the LHC is an 
injector, and the research done there may benefit the VHEeP project.  Considering a faster refill time, but with a less "clean" 
proton bunch, may be an option~\cite{private:burt}.  Another method to increase the effective luminosity is to have more than one 
$ep$ interaction point along the plasma accelerator, using the natural spacing of the bunches so that an electron or proton bunch 
will be brought into collision at multiple points.  With a small proton bunch spacing, e.g. 5\,ns, multiple collisions per proton bunch, 
say 10, and with each electron bunch, could occur within a short distance, thereby requiring only one detector, but 
increasing the luminosity by a factor of 10.

Obtaining higher, but realistic, luminosities is an issue that needs significant and further investigation.  This design assumes LHC 
bunches as currently used; if these can be shortened and the plasma density correspondingly increased, higher accelerating gradients 
of multi-GV/m could be achieved.  This could lead to a shorter acceleration section.  Overall, more detailed studies 
are needed to develop the basic and simple accelerator concept in Fig.~\ref{fig:acc} into a technical concept.

\section{The physics case}

Given the high beam energies, VHEeP will be able to access very low Bjorken $x$ and, with sufficient luminosity, high virtualities, 
$Q^2$.  A $Q^2$ value of 1\,GeV$^2$, a region readily measured at HERA,  corresponds to an $x$ value of $10^{-8}$.  It should 
be noted that the lowest value of $Q^2$ measured at HERA was $Q^2 = 0.045$\,GeV$^2$, which at VHEeP corresponds to a 
minimum $x$ value of $5 \times 10^{-10}$.  At these low $Q^2$ values, a significant number of events is expected.  The detector 
design will need to cope with this large range in kinematics.  The polar angle of produced hadrons is distributed over 0$^\circ$, 
corresponding to the proton direction, to 
180$^\circ$, but with a peak at 0$^\circ$ and even stronger peak at 180$^\circ$.  The events at low angles are due to events at 
high $x$, whereas the events with hadrons at high angles are dominated by events at low $x$.  A central detector is needed as 
well as instrumentation close to the beamline to measure both electrons and hadrons.  Further and more detailed studies of the 
detector design are needed.

The energy dependence of hadronic cross sections are poorly understood.  Predictions calculated from first principles are often 
not available and so phenomenological models are used to describe the dependence.  Being able to measure the energy 
dependence, particularly with the long lever arm presented by VHEeP, will deepen our understanding of QCD and the structure 
of matter.  Assuming measurements from VHEeP of the total $\gamma p$ cross section up to $W \sim 6$\,TeV, even with very low 
luminosities, the data will be able to strongly constrain the energy dependence of the total cross section and hence provide a clearer 
picture of QCD.  It should be noted  that the multi-TeV energies and hence large lever arm attainable at VHEeP are necessary to do this.
A photon--proton collision of $W = 6$\,TeV, corresponding to photon and proton energies of, respectively, 1.3\,TeV and 7\,TeV, 
is equivalent to a 20\,PeV photon on a fixed target.  This extends significantly 
into the region of ultra high energy cosmic rays.  Therefore VHEeP data could be used to constrain cosmic-ray air-shower 
simulations and so will be of benefit to understanding the nature of cosmic rays at the highest energies.
In a similar vain, vector meson production is dependent on the partonic distributions in the proton and given the need for two gluons 
from the proton to create a vector meson, is particularly sensitive to saturation of the parton densities or 
other effects.  Data from VHEeP will be able to determine the behaviour of the cross sections, which must take on another form and 
start to level out, be it through saturation or some other mechanism.
Also for scattering cross sections for virtual photons on protons, a change of the energy dependence is  expected to 
become visible in the VHEeP kinematic range.  This should yield exciting and unique information on the fundamental underlying physics 
at the heart of the high energy dependence of hadronic cross sections.

At such a high centre-of-mass energy, VHEeP also has sensitivity to physics beyond the Standard Model at high $Q^2$.  Its particular 
strength is the search for leptoquarks which would be produced resonantly.  The sensitivity is up to the kinematic limit of the 
centre-of-mass energy, well beyond the limits from HERA and also beyond those from the LHC experiments.

In addition, standard tests of QCD will be possible in a new kinematic regime as well as running at different electron beam energies and 
electron--ion scattering.  Investigation of the possibilities will be pursued.

\section{Outlook}

An idea for a very high energy electron--proton collider has been presented, focusing in particular on the particle physics case for such a 
machine with a centre-of-mass energy of 9\,TeV.  With this significant extension in kinematic reach, QCD and the structure of matter 
will be probed in a region where it is not understood.  Further work is needed and will be pursued to strengthen and extend the physics 
case of an $ep$ collider at such high energies.  In parallel, investigations of the technical feasibility of the accelerator system and its 
ultimate performance are needed as well as a more detailed detector design.

\end{document}